\documentclass{elsart}

\usepackage{color}
\usepackage{times,amssymb, amsmath}

\usepackage{graphicx}

\newcommand{\ora}{\overrightarrow}

\begin{document}

\DeclareGraphicsExtensions{.eps}

\begin{frontmatter}

\title{Position Reconstruction in Drift Chambers operated with Xe,CO$_2$(15\%)}

\author[B]{C. Adler},
\author[A]{A. Andronic},
\author[A]{H. Appelsh\"{a}user},
\author[B]{J. Bielcikova},
\author[H]{C. Blume},
\author[A]{P. Braun-Munzinger},
\author[C]{D. Bucher},
\author[A]{O. Busch},
\author[E,B]{V. C\u{a}t\u{a}nescu},
\author[E,A]{M. Ciobanu},
\author[A]{H. Daues},
\author[B]{D. Emschermann},
\author[F]{O. Fateev},
\author[A]{P. Foka},
\author[A]{C. Garabatos},
\author[G]{T. Gunji},
\author[B]{N. Herrmann},
\author[G]{M. Inuzuka},
\author[B,I]{M. Ivanov},
\author[F]{E. Kislov},
\author[D]{V. Lindenstruth},
\author[A]{C. Lippmann\corauthref{cor1}}
\corauth[cor1]{Corresponding author},
\ead{C.Lippmann@gsi.de}
\ead[url]{http://www-linux.gsi.de/{\~{}}lippmann}
\author[B]{W. Ludolphs},
\author[B]{T. Mahmoud},
\author[B]{V. Petracek},
\author[E]{M. Petrovici},
\author[A]{S. Radomski},
\author[B]{I. Rusanov},
\author[A]{A. Sandoval},
\author[C]{R. Santo},
\author[B]{R. Schicker},
\author[A]{K. Schwarz},
\author[A]{R.S. Simon},
\author[F]{L. Smykov},
\author[B]{H.K. Soltveit},
\author[B]{J. Stachel},
\author[A]{H. Stelzer},
\author[A]{G. Tsiledakis},
\author[B]{B. Vulpescu},
\author[C]{J.P. Wessels},
\author[B]{B. Windelband},
\author[F]{V. Yurevich},
\author[F]{Yu. Zanevsky},
\author[C]{O. Zaudtke}

for the ALICE collaboration.

\address[A]{Gesellschaft f\"{u}r Schwerionenforschung, Darmstadt, Germany}
\address[B]{Physikalisches Institut der Universit{\"a}t Heidelberg, Germany}
\address[C]{Institut f\"{u}r Kernphysik, Universit{\"a}t M{\"u}nster, Germany}
\address[D]{Kirchhoff-Institut f\"{u}r Physik, Heidelberg, Germany}
\address[E]{NIPNE Bucharest, Romania}
\address[F]{JINR Dubna, Russia}
\address[G]{University of Tokyo, Japan}
\address[H]{Institut f\"{u}r Kernphysik, Universit{\"a}t Frankfurt, Germany}
\address[I]{PH Division, CERN, Switzerland}


\begin{abstract}

  We present measurements of position and angular resolution of drift
  chambers operated with a Xe,CO$_2$(15\%) mixture. The results are
  compared to Monte Carlo simulations and important systematic effects
  -- in particular the dispersive nature of the absorption of transition
  radiation and non-linearities -- are discussed. The measurements were
  carried out with prototype drift chambers of the ALICE Transition
  Radiation Detector, but our findings can be generalized to other
  drift chambers with similar geometry, where the electron drift is
  perpendicular to the wire planes.

\end{abstract}


\end{frontmatter}


\section{Introduction}
\label{Intro}

Around 40 years after their introduction \cite{charpak1}, multiwire
proportional chambers (MWPCs) and drift chambers are widely in use
in particle physics experiments and other fields. The main
properties of these detectors, i.e. good position, timing and energy
resolution and competitive rate capabilities at low cost, make them
very attractive for usage in large scale high-energy physics and
heavy ion experiments.

The characteristics of these detectors have been extensively studied
in the past \cite{blum}. However, with the stringent requirements of
modern experiments and with new applications for proportional
chambers, still a large effort is devoted to the understanding and
improvement of existing designs and to the development of new
concepts.

In this publication, we investigate the position reconstruction
capabilities of the Transition Radiation Detector (TRD) \cite{TRD}
of the ALICE
experiment. ALICE\footnote{A Large Ion Collider Experiment.} is a
dedicated heavy ion experiment to be operated at the Large Hadron
Collider (LHC) at CERN. The ALICE TRD offers three dimensional
tracking, electron/pion identification and -- combining these two
capabilities -- a fast trigger on high-$p_t$ electrons and jets. At
the very high particle multiplicities anticipated in central Pb-Pb
collisions (several thousand charged particles per unit of rapidity
at mid-rapidity) at the LHC, these are very ambitious tasks. To be
able to select stiff electron tracks, an excellent position
reconstruction performance in the bending plane of the ALICE
magnetic field is required, characterized by a position resolution
below 400\,$\mu$m and an angular resolution better than 1$\,^\circ$.

\section{The ALICE TRD}
\label{alicetrd}

In this section we describe the transition radiation detector of the
ALICE experiment, in particular its position reconstruction and
particle identification capabilities.

\subsection{General Description and Working Principle}
\label{Principle}

\begin{figure}[t]
  \begin{center}
    \includegraphics[width=13cm]{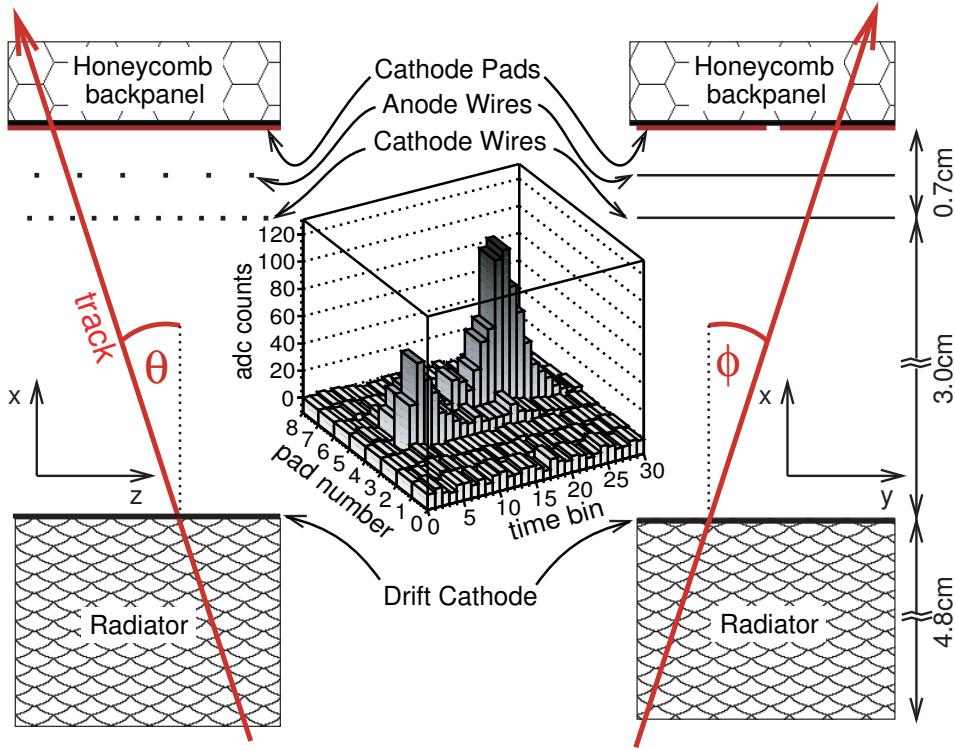}
    \caption{Schematic view of a TRD chamber (not to scale). The left
    cross section shows a projection of the chamber in the $x$-$z$ plane,
    perpendicular to the wires, the right one shows a projection in
    the $x$-$y$ plane, which is the bending plane of the particles in the
    ALICE magnetic field. A particle trajectory is also sketched. The
    insert shows the pulse height versus drift time on eight cathode
    pads for an example event. One time bin corresponds to 100\,ns.}
    \label{xsec}
    \hspace{5mm}
  \end{center}
\end{figure}

The ALICE TRD consists of 540 chambers surrounding the Time
Projection Chamber (TPC) in six layers at an overall length of about
7\,m. The total sensitive area is roughly 750\,m$^2$; the largest
chamber is 159\,cm long and 120\,cm wide. Each module is about
13\,cm thick, including radiator, electronics and cooling. The total
anticipated radiation thickness for six layers is about $0.15 X_0$.

A schematic cross section of a TRD module is shown in Fig.
\ref{xsec}. The gas volume is subdivided into a 3\,cm drift region
and a 0.7\,cm amplification region, separated by a cathode wire grid
with 0.25\,cm wire pitch and 75\,$\mu$m wire diameter. The anode
wires have 0.5\,cm pitch and 20\,$\mu$m diameter. The drift chambers
are equipped with cathode pads of varying sizes\footnote{The width
of the pads ranges from 0.664 to 0.818\,cm, their length from 7.5 to
9\,cm.} and are read out via charge sensitive preamplifiers/shapers
(PASA). The whole system will consist of about 1.18 million channels
(readout pads). The maximum drift time is about 2\,$\mu$s and the
induced signal is sampled on all channels at 10\,MHz to record the
time evolution of the signal \cite{signal,gain}. A typical signal
generated by a particle track through a prototype drift chamber is
also shown in Fig. \ref{xsec}.

A 4.8\,cm thick radiator is placed in front of each gas volume. This
radiator is a sandwich of polypropylene fibers and Rohacell foam,
which provides many interfaces between materials with different
dielectric constants. Transition radiation (TR) is emitted by
particles traversing the radiator with a velocity larger than a
certain threshold \cite{TR}, which for typical materials corresponds
to a Lorentz factor of $\gamma \approx 1000$. The produced TR
photons have energies in the X-ray range (1 to 30\,keV)
\cite{TRspec} and a high-Z gas mixture (Xe, CO$_2$ (15\,\%)) is used
to provide efficient absorption of these photons.

\subsection{Electron Identification}
\label{eid}

The TRD will provide electron identification for momenta above
1\,GeV/c \cite{pioeff}. To discriminate electrons from the large
background of pions two characteristic phenomena are used:

\begin{itemize}
\item[i)] The ionization energy loss \cite{dedx} at the momentum region
  of interest is larger for electrons than for pions, since here
  electrons are at the plateau of ionization energy loss, while pions
  are minimum ionizing or on the relativistic rise.
\item[ii)] In the momentum range considered, only electrons exceed the
  TR production threshold.
\end{itemize}

\begin{figure}[t]
  \begin{center}
    \includegraphics[width=9cm]{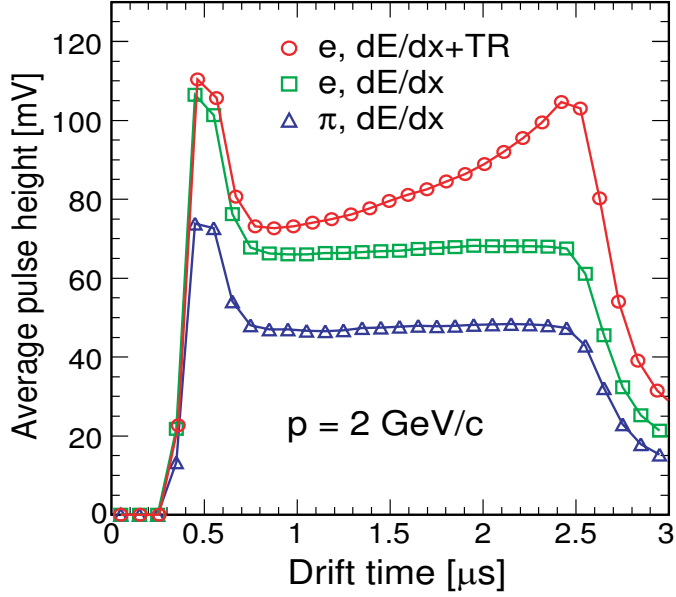}
    \caption{Measured average pulse height as a function of the drift
    time for pions and electrons (with and without radiator). The peak
    at short drift times is due to the fact that electrons produced
    by ionization in the amplification region drift towards the anode
    wires from both sides of the wire plane, which leads to an
    approximate doubling of the average pulse height. In general, the
    average pulse height is larger for electrons. TR adds a significant
    energy deposit and introduces the characteristic signal shape of
    electrons, determined by the exponential absorption probability
    distribution of TR photons in the gas.}
    \label{signals}
   \hspace{5mm}
  \end{center}
\end{figure}

Fig. \ref{signals} shows the mean pulse height as a function of the
drift time for pions and electrons \cite{signal}. Here, and in the
following, the time zero is arbitrarily shifted to facilitate a
simultaneous measurement of the baseline and of noise. Due to the
larger ionization energy loss at these specific conditions
($p=2$\,GeV/c) the mean signal is about 40\,\% larger in the case of
electrons (without radiators). With radiators the energy deposited
by absorbed TR photons contributes considerably to the mean
amplitude of the electrons. The characteristic signal shape for
electrons with radiators is determined by the exponential
probability distribution for the absorption of TR photons in the gas
mixture.

\subsection{Tracking}
\label{tracking}

In this publication, we focus on the position reconstruction
performance of the ALICE TRD in the bending plane of the particles
in the ALICE magnetic field, which is parallel to the wires of the
TRD and to the electric drift field. This defines the transverse
momentum resolution of the TRD. In the third dimension, parallel to
the magnetic field lines, the resolution is limited by larger pads
and by the discrete wire positions. A tilted-pad design will be
employed to increase the tracking capabilities in this direction.

\begin{figure}[ht]
  \begin{center}
    \includegraphics[width=9cm]{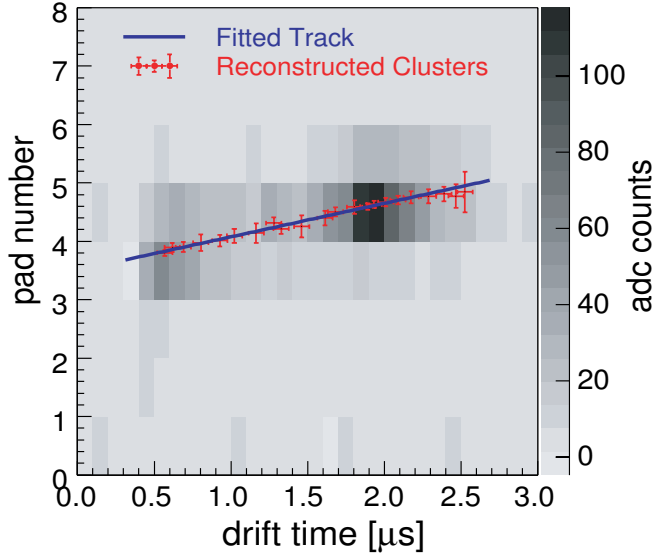}
    \caption{Same example event as in Fig. \ref{xsec}. The
    reconstructed clusters and a fitted track are overlayed.}
    \label{event2}
   \hspace{5mm}
   \end{center}
\end{figure}

An example event in a TRD chamber is shown in Fig. \ref{event2}.
From the pulse height distribution on eight adjacent pads of
$0.75$\,cm width, the cluster position is reconstructed as a
function of the drift time. In this context a cluster represents
electrons triggering avalanches in a given time bin. For a
discussion of the position reconstruction method see section
\ref{Reconstr}. Electrons generated close to the anode wire grid
have a small drift time and induce signals corresponding to a small
time bin number. Electrons originating close to the drift electrode,
on the other hand, have a larger drift time and thus correspond to
larger time bin numbers. The drift time can be translated to a
position (distance from the anode wire plane) if the drift velocity
is known. For the ALICE TRD we aim for a drift velocity of around
1.5\,cm/$\mu$s in the drift region. When the clusters at all time
bins are reconstructed, a straight line fit -- as in Fig.
\ref{event2} -- defines the reconstructed track. The reconstructed
angle $\phi_{rec}$ is obtained by:

\begin{equation}
  \tan \phi_{rec} \ = \ \frac{a\,W}{v_D^{av}} \ ,
  \label{trackvd}
\end{equation}

where $a$ is the slope parameter of the linear fit line in pad
units, $W$ is the pad width in cm and $v_D^{av}$ is the average
drift velocity of electrons in the detector in cm/$\mu$s.

\section{Experimental Setup}
\label{ExpSetup}

The measurements were carried out at momenta of 1 to 6\,GeV/c at the
T10 secondary beam line at the CERN PS. A sketch of the beam setup
is shown in \cite{obusch}. The beam was a mixture of electrons and
negative pions with a momentum spread of about 1\,\%. Clean samples
of each particle type were selected using coincident thresholds on
two \u{C}erenkov detectors and a lead-glass calorimeter. Position
reference was provided by two silicon detectors with a 50\,$\mu$m
strip pitch. With this silicon telescope the beam divergence was
found to be around 0.1$\,^\circ$ ($\sigma$).

We tested four identical prototype drift
chambers\footnote{Generally, in this publication we will average
  over the behaviour of these four chambers, thus increasing the
  statistics of the measurements. Only where the performance of
  the four chambers is expected to be different, e.g. due to the
  track curvature in the magnetic field, we show results for a
single chamber.} with a construction similar to that for the final
TRD, but with a smaller active area (25\,$\times$\,32\,cm$^2$). The
dimensions of the pads were 0.75\,$\times$\,8\,cm$^2$. We used a
prototype of the PASA with a noise on-detector of about 1000
electrons (r.m.s.). The FWHM of the output pulse is about 100\,ns
for an input step function. The nominal gain of the PASA is
12\,mV/fC but during the present measurements we used a gain of
6\,mV/fC to better match to the range of the employed Flash ADC
system with 0.6\,V voltage swing. The high voltage at the anode
wires was adjusted to four values corresponding to gas gains of
2400, 3900, 6200 and 9600. A gain value of around 4000 is
anticipated as the nominal value for the ALICE TRD.

\section{Detector Simulations}
\label{Sim}

For simulations of the TRD performance we use AliRoot
\cite{Aliroot}, the ALICE software package. AliRoot provides an
object oriented framework for event simulations and reconstruction
in the ALICE detector. The TRD part of AliRoot contains a full
microscopic simulation of the detector physics processes. The
interaction of the charged particles with the detector materials and
their energy loss is simulated using Geant 3.21 \cite{Geant}. Since
the production of transition radiation is not included in Geant 3,
it was explicitly added to AliRoot. We use a momentum dependent
parameterization which applies an approximate formula for the TR
yield of a regular stack of foils with fixed thickness, including
absorption \cite{TRD,Fabjan}.

The energy transfers in primary collisions and the energy deposited
by TR are converted into a number of secondary electrons and the
electron collection is simulated taking into account electron drift
and diffusion, amplification fluctuations, the distribution of the
induced charge on the cathode pads (pad response), the time response
of the detector (ion tail) as well as that of the electronics, and
finally noise. The deposited charge is translated into raw-data-like
ADC signals which then serve as input for track reconstruction.

In this publication we use AliRoot to study in some detail the
different contributions to the position reconstruction performance
of the ALICE TRD. During our studies some changes had to be made in
the AliRoot code, which will be described in the following.

\subsection{Transition Radiation Absorption} \label{TRabsor}

TR photons are emitted in the radiator with an angular distribution
about the direction of the emitting particle, which is sharply
peaked at $1/\gamma$ \cite{TR}. As a consequence, the TR photons
cannot be separated from the incident electron track and contribute
to the tracking information. The drift chamber detects the
photoelectron ejected from a gas atom and the charge that is
released by the secondary processes. This can introduce a
considerable smearing of charge deposit and hence a degradation of
the tracking performance of the detector\footnote{The physical
limitations imposed on the imaging quality of a xenon-filled MWPC
X-ray imaging detector are studied in detail in \cite{bateman}.}.

In the following, we only consider the xenon atoms; photon
interactions with CO$_2$ atoms are neglected. In the absorption
process, a photoelectron of energy $E_e=E_X-E_S$ is created, where
$E_X$ is the energy of the TR photon and $E_S$ is the binding energy
of the photoelectron\footnote{The xenon K-shell binding energy
 is about 35\,keV, the average L- and M-shell binding energies are
 about 5.1 and 0.9\,keV, respectively.}. At the relevant X-ray
energies, the photoelectron is emitted preferentially in a plane
perpendicular to the incoming photon track \cite{charpak2}. Even
though subsequent multiple scattering and ionizing collisions with
the gas molecules randomize the photoelectron trajectory, the charge
will be deposited some distance away from the track. The practical
range $R(E_e)$ of this electron in a gas can be calculated according
to \cite{blum}

\begin{equation}
  R(E_e) \ = \ A\,E_e\,\left(1-\frac{B}{1+C\,E_e}\right) \quad
  {\bf \frac{g}{cm^2}} \ ,
\label{range}
\end{equation}

where $A = 5.37\cdot 10^{-4}$\, g\,cm$^{-2}$\,keV$^{-1}$, $B=0.9815$
and $C = 3.123\cdot 10^{-3}$\,keV$^{-1}$. As an example, for our gas
mixture the range of a 10\,keV electron is about 500\,$\mu$m.

The emission of the photoelectron leaves a hole in the shell which
will be filled with electrons from higher shells. This de-excitation
occurs by emission of either an Auger electron or a fluorescence
photon. The probability for de-excitation by photon emission is
determined by the fluorescence yield, which is 0.87 for the xenon
K-shell \cite{sauli}. The fluorescence photon energy is $E_S-E_T$,
where $E_T$ is the binding energy of the second shell involved in
the transition. The emission of fluorescence photons is isotropic
and their absorption length in the gas is exponentially distributed
with a mean that is given by the attenuation coefficient. As an
example, a K-shell fluorescence photon will have an energy of $E_K
\approx 35$\,keV; in pure xenon it will have an absorption length of
24\,cm and can generate a background hit at a distant position.
However, since the energy of the largest part of the TR photons is
below $E_K$, these are rare events (i.e. 1.4\% at $p=2$\,GeV/c).
L-shell fluorescence X-rays, on the other hand, are very common (due
to the mean TR photon energies around 10\,keV and the large
fluorescence yield); they carry an energy of around 5\,keV and have
an absorption length around 0.4\,cm in Xenon. Auger electron
emission is also isotropic. The range of the Auger electrons is
calculated by Eq. \ref{range}. As an example, an Auger electron
emerging from the L-shell ($E_K-2E_L\approx 25$\,keV) has a range
around 0.2\,cm in Xenon. A simplified picture of the just described
secondary processes has been added to the AliRoot code to allow
studies of their influence on the position reconstruction
performance of the TRD.

\subsection{Electron Drift Path}
\label{NonIso}

In drift chambers one takes advantage of a unique relation between
the position of primary ionization electrons ($x$) and the drift
time ($t_D$) to the nearest anode wire, where the electrons generate
avalanches. To precisely reconstruct the position of the passage of
the particle through the detector one generally wants to know this
space-time relation, which may not be linear:

\begin{equation}\label{driftl}
    x \ = \ \int\limits_0^{t_D}\,v_D(t)\,dt \ .
\end{equation}

\begin{figure}[t]
  \begin{center}
    \includegraphics[width=9cm]{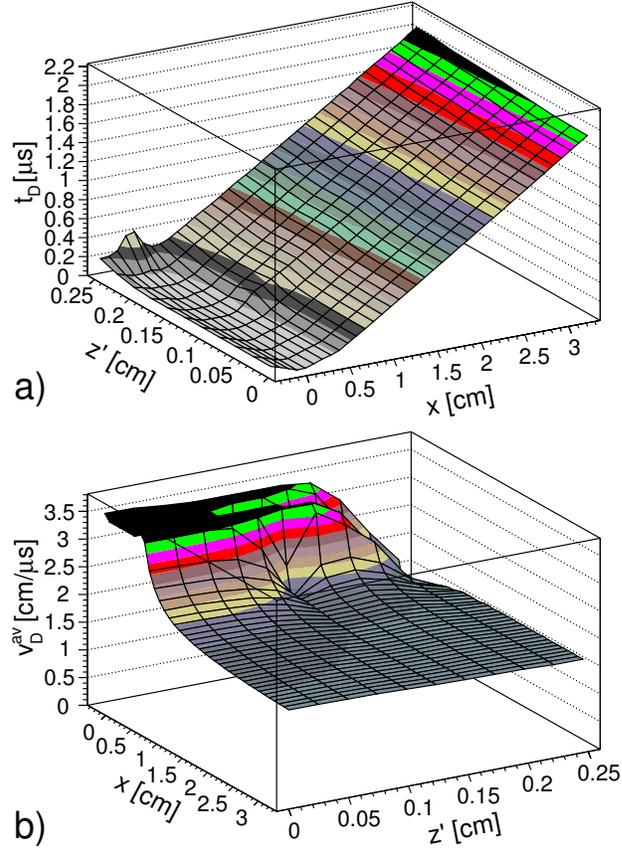}
    \caption{(a) Average drift time $t_D$ for electrons at different
    positions in the drift volume. The anode wire is situated at
    ($x=0$\,cm, $z'=0$\,cm). $z'$ is the distance to the nearest anode
    wire in $z$ (Figs. \ref{xsec},\ref{drlines}). The calculation was
    done with GARFIELD for the Xe,CO$_2$ (15\,\%) gas mixture, an anode
    wire voltage of 1550\,V and a drift voltage of -1950\,V.
    (b) Corresponding distribution of the average drift velocity.
    Note that the coordinate system has been rotated as compared to
    (a) for better visibility.}
    \label{noniso}
    \hspace{5mm}
  \end{center}
\end{figure}

Here $v_D(t)$ is the local drift velocity at time $t$. For constant
drift velocity the space-time relation Eq. \ref{driftl} becomes linear:

\begin{equation}\label{driftl2}
    x \ = \ v_D\,t_D\ .
\end{equation}

In the TRD $v_D$ is constant in a large fraction of the detector
($v_D=v_0=1.5$\,cm/$\mu$s in the drift region), but (in general)
higher in the amplification region. However, one can approximate:

\begin{equation}\label{driftl3}
    x \ \approx \ v_D^{av}\,t_D\ ,
\end{equation}

where $v_D^{av}$ is an average drift velocity. We used GARFIELD
\cite{garfield} to calculate drift times $t_D$ (Fig. \ref{noniso}a)
and average drift velocities $v_D^{av}$ (Fig. \ref{noniso}b) for
electrons generated at a given ($x$, $z'$) position. Here $z'$
denotes the lateral distance of the position of the drifting
electrons to the closest anode wire ($0\leq z' \leq 0.25$\,cm). A
large number of electrons were drifted from each position (including
diffusion) and the most probable drift time was taken. We find that
$v_D^{av}$ is approximately equal to $v_0$ only close to the drift cathode
($x \approx 3.35$\,cm), but increases as $x$ decreases. In the
amplification region it is in general more than twice as large
(3.5\,cm/$\mu$s).

\begin{figure}[t]
  \begin{center}
    \includegraphics[width=9cm]{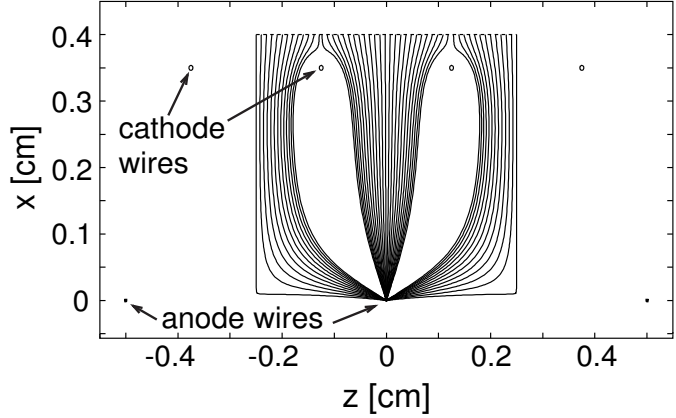}
    \caption{Ideal drift lines for electrons coming from the drift region
    (from the top) at different $z$ positions. The anode wire plane is
    at $x = 0$\,cm; the cathode wire plane is at $x = 0.35$\,cm.}
    \label{drlines}
    \hspace{5mm}
  \end{center}
\end{figure}

However, the drift time $t_D$ depends also on $z'$. Fig.
\ref{drlines} shows the different drift paths for electrons coming
from the drift region at different $z$-positions. At a given
distance from the anode wire plane (which is situated at $x=0$\,cm),
the shortest drift time is given by electrons drifting at
$z'=0$\,cm. Electrons drifting at $z'=0.25$\,cm have a longer drift
path and -- on top of that -- cross the low field region between two
anode wires. Consequently, we observe in Fig. \ref{noniso}a a drift
time offset depending on $z'$. For electrons coming from the drift
cathode ($x \approx 3.35$\,cm) it is around 120\,ns as compared to
the value at ($x=0$\,cm, $z'\approx 0$\,cm). For electrons
originating in the low field region at ($x=0$\,cm, $z'=0.25$\,cm) we
find $v_D^{av}<1$\,cm/$\mu$s, resulting in a drift time
offset\footnote{Diffusion considerably broadens the drift time
distributions, especially in this region. Fig. \ref{noniso} only
shows the most probable drift times.} up to 430\,ns!

The non-linearity of the space-time relationship as illustrated in
Fig. \ref{noniso} has been added to the AliRoot code.

\subsection{Pad Response Function}
\label{prf}

Proportional chambers often feature a cathode plane subdivided into
separate strips or -- like the ALICE TRD -- pads with independent
charge sensitive readout for the purpose of localizing the avalanche
with a precision that is a fraction of the strip or pad width $W$. A
parameter that strongly influences the distribution of the induced
charge on the cathode plane is the angular position of the avalanche
at the anode wire \cite{mathieson1}. However, in most practical
situations it is not possible or desirable to restrict or control
this quantity so that one generally observes a resultant effect due
to all avalanche angles. There exists an empirical formula for the
induced charge distribution $\rho(y)$ by Mathieson \cite{mathieson2}
that describes well such average behaviour in symmetric MWPCs along
the anode wires\footnote{A symmetric MWPC consists of a plane of
anode wires centered between two planar cathodes.}. The coordinate
$y$ is given by the wire direction (Fig. \ref{xsec}).

The pad response function (PRF) $P(y)$ is obtained by integration
of $\rho(y)$ over the width of the strip or pad

\begin{equation}
  P(y) \ = \ \int\limits_{y-W/2}^{y+W/2}\rho(y')\,\d y'\ .
  \label{prfcalc}
\end{equation}

The ALICE TRD however is not exactly a MWPC, but is extended by a
drift volume that is separated from the amplification volume by a
cathode wire grid. As we shall see in this section, the Mathieson
formula can nevertheless be used to calculate $P(y)$ for this design
to a rather good accuracy.

The exact PRF $P(y)$ can be obtained by employing the weighting
field formalism. The weighting field $\ora{E}_W(y)$ is the
(imagined) electric field in the detector when the readout electrode
is set to 1\,V while all other electrodes are grounded. The field
$\ora{E}_W(y)$ is generally used to calculate induced currents in
arbitrary electrode geometries, using the Ramo theorem\footnote{Also
known as the reciprocity theorem.} \cite{ramo} via

\begin{equation}
  i(t) \ = \ - q\ \ora{E}_W(\ora{r}(t))\cdot\ora{v}(t) \ .
\end{equation}

The current $i$ that is induced on a readout electrode at time $t$
by a charge $q$ moving with velocity $\ora{v}$ is proportional to
the weighting field $\ora{E}_W$ at the position $\ora{r}$ of the
charge. Calculating $\ora{E}_W$ as a function of $y$ on the cathode
plane yields the cathode charge distribution $\rho(y)$ for a given
geometry. From this we can then calculate $P(y)$ following Eq.
\ref{prfcalc}.

\begin{figure}[t]
  \begin{center}
    \includegraphics[width=9cm]{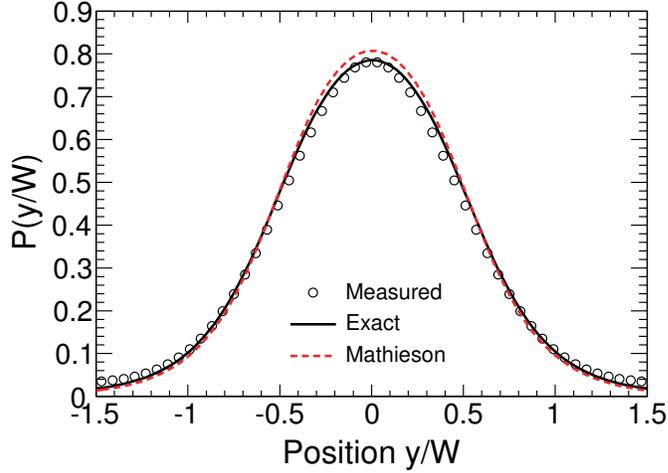}
    \caption{Pad response functions for $W = 0.75$\,cm. The circles
    show the measured PRF, the dashed line shows the results of a
    calculation using the Mathieson formula and the solid line
    shows the exact calculation for the ALICE TRD geometry.}
    \label{prfpl}
    \hspace{5mm}
  \end{center}
\end{figure}

We used GARFIELD to calculate $\rho(y)$ for the ALICE TRD geometry
(Fig. \ref{prfpl}). Since we are -- as already mentioned -- not
interested in the influence of the angular position of the avalanche
around the anode wire, we average over many angles.

The exact method confirms that the Mathieson formula is a good
approximation. In general, the PRF can be well approximated by a
Gaussian curve \cite{blum}. The standard deviations $\sigma_P$ of
Gaussian fits to the $P(y)$ (in pad units) are 0.495 and 0.496 for
the measured PRF and the exact calculation for the ALICE TRD
geometry, respectively. For the Mathieson formula and for the exact
calculation for a MWPC geometry with similar parameters\footnote{No
drift region.} we find 0.482 and 0.485, respectively. The PRF for a
symmetric MWPC is narrower by about 3\,\% (in $\sigma_P$) as
compared to the TRD geometry, for our specific wire diameters, wire
pitches and anode-cathode separation. The PRFs calculated with
GARFIELD, as described in this section, are used for the simulation
of the pad response in AliRoot.

\section{Position Reconstruction and Systematic Effects}
\label{SysEff}

In this section we present some first results on the position
reconstruction performance of the ALICE TRD and describe the
different systematic effects.

\subsection{Definitions of Resolutions}
\label{Defs}

The residuals for a given track are defined as the distance between the
position of the reconstructed cluster $(y_t)_{cl}$ and the position
of the reconstructed track $(y_t)_{fit}$ for each time bin $t$:

\begin{equation}\label{residuals}
    \Delta_y \ = \ (y_t)_{cl} \ - \ (y_t)_{fit}\ .
\end{equation}

As the position resolution $\sigma_y$ of the detector we define the
sigma of a Gaussian fit (within $3\sigma$) to the distribution of
residuals $\Delta_y$ for a large number of tracks. This resolution
does not depend on 'external effects' like multiple scattering in
front of the gas volume of the drift chamber and/or beam divergence.
It thus represents the detector-intrinsic position resolution. As
the angular resolution we define the width $\sigma_\phi$ of a
Gaussian fit (within 3$\sigma$) to the distributions of the
reconstructed angles. This resolution includes the mentioned
external effects.

\subsection{Tail Cancellation}
\label{TC}

\begin{figure}[t]
  \begin{center}
    \includegraphics[width=9cm]{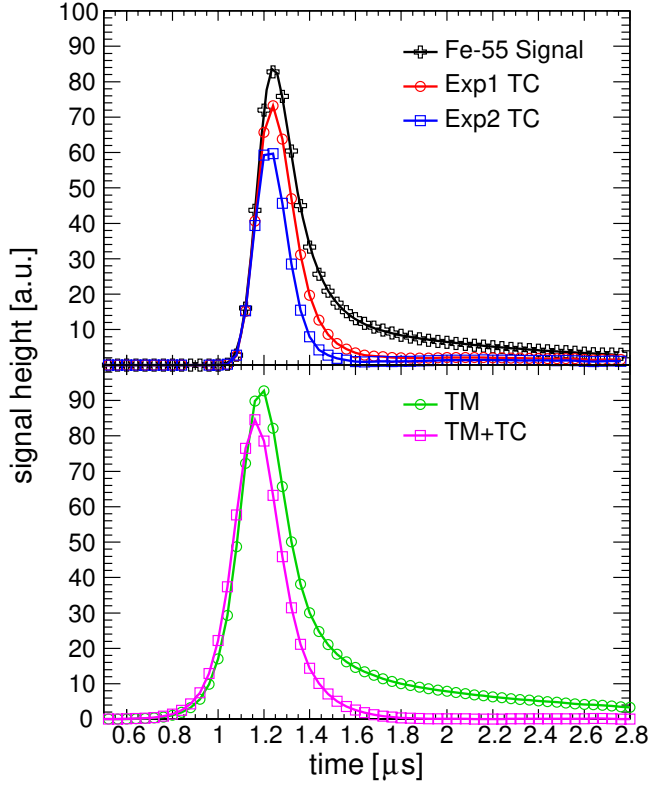}
    \caption{Time dependence of the average PASA pulse height for
    $^{55}$Fe X-rays. Due to the pointlike charge deposit by
    X-rays this signal is almost identical to the time response
    function (TRF). In the shown signal diffusion is included, in the
    TRF not. The upper panel shows the original signal and the
    signal after applying a tail cancellation with one and two
    exponential functions (Exp1TC, Exp2TC) . The lower panel shows
    the effect of adding the short tail component to the left
    (tail making, TM), and of subsequent tail cancellation
    (long component).}
    \label{fe55}
    \hspace{5mm}
  \end{center}
\end{figure}

The signals that are read out from the cathode pads are induced by
the positive ions generated in the electron avalanches near the
anode wires. Since the massive ions move slowly compared to the
electrons, the signals exhibit long tails. Convolution with the
response of the PASA yields the time response function (TRF), which
is asymmetric. For our specific chamber geometry and electronics,
the tail of the TRF can be well approximated by the sum of two
exponential functions with characteristic decay times

\begin{equation}
  T_{short} \approx 0.10\,{\bf \mu\mbox{s}} \quad\mbox{and}\quad
  T_{long}  \approx 0.93\,{\bf \mu\mbox{s}}\ .
\end{equation}

The TRF gives rise to a strong correlation between the signal
amplitude in subsequent time bins. This is in general a problem also
in other related detectors, in particular in TPCs, since it biases
the position measurement results as a function of time. In the case
of the TRD the correlations affect especially the angle measurement
(see Fig. \ref{event2}). A way to minimize the effect is to remove
the tails from the data by deconvolution (tail cancellation). Three
different methods are studied here:

\begin{itemize}
  \item The one exponential tail cancellation (Exp1TC) subtracts
    the tail (for each time bin) as a function of time. Here the
    tail is assumed to be a one exponential function with decay
    time $T_{long}$.
  \item The two exponential tail cancellation (Exp2TC) subtracts
    accordingly a tail that is assumed to consist of a superposition
    two exponential functions with decay times  $T_{long}$ and
    $T_{short}$.
  \item Finally we also apply a signal symmetrization (TM+TC)
    that first replicates the tail with $T_{short}$ at the times
    preceding the maximum (tail maker, TM) and then subtracts only
    the long component similar to the first mentioned method (TC).
\end{itemize}

The effects of the three different methods on the TRF are shown in
Fig. \ref{fe55}. The tail of the original TRF is largely reduced  by
the Exp1TC method and the maximum signal amplitude is lowered by
around 10\%. However, the TRF is not fully symmetrized, so we expect
some correlation to remain, if this method is used. The Exp2TC
method symmetrizes the TRF but the effective signal amplitude is
reduced by around 30\,\%, which introduces a considerable
degradation in the signal-to-noise ratio. The TM+TC method also
symmetrizes the TRF but without the drawback of a reduction in
signal amplitude.

\begin{figure}[pt]
  \begin{center}
    \includegraphics[width=12cm]{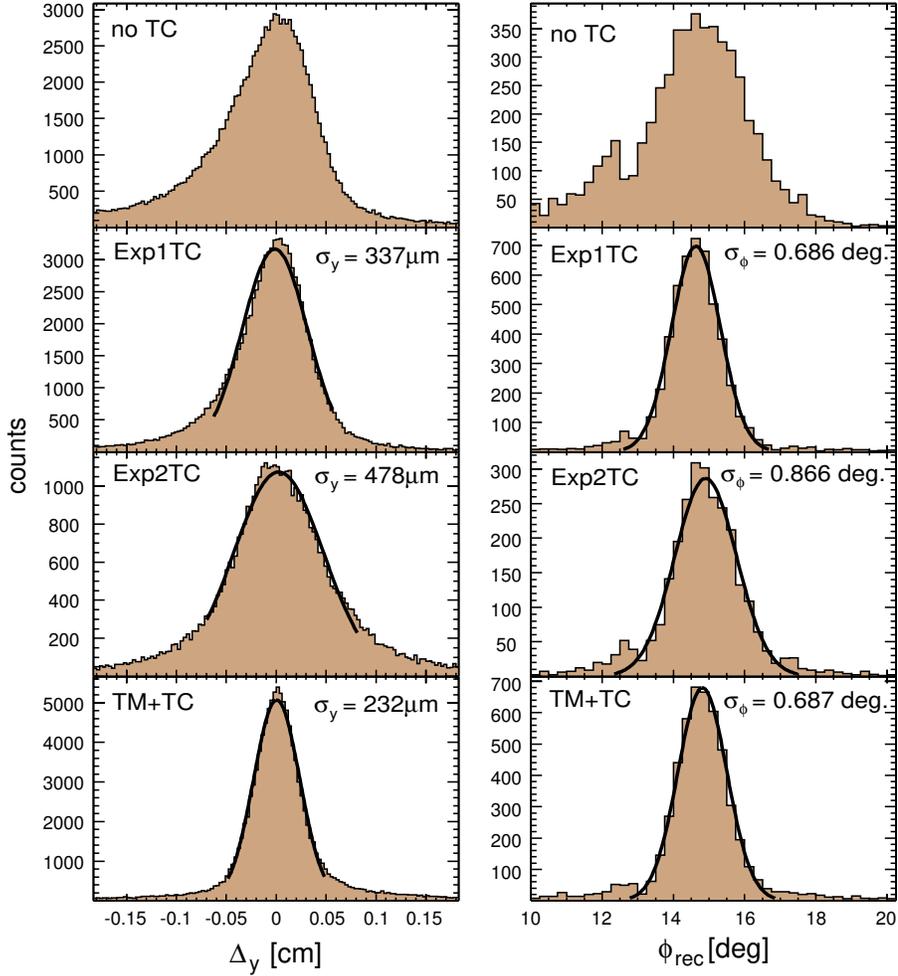}
    \caption{Measured histograms of residuals $\Delta_y$ (left panels)
    and of the reconstructed angles (right panels) for 6\,GeV/c
    pions. We show data without tail cancellation and for the Exp1TC,
    Exp2TC and TM+TC methods (from top to bottom). Where applicable,
    Gaussian fits are also shown.}
    \hspace{5mm}
    \label{reshist}
  \end{center}
\end{figure}

Fig. \ref{reshist} shows example histograms of the residuals
$\Delta_y$ of fitted tracks and of the reconstructed angles for the
different tail cancellation methods described in section \ref{TC}.
The incident angle of the beam was $\phi \approx 15\,^\circ$. If no
tail cancellation is applied, the distributions are very broad, with
pronounced tails as a result of the mentioned correlations. Since an
entry at a given time bin increases the amplitude at later time bins
on the same pad through the TRF, the reconstructed angles are
generally shifted towards smaller values by this effect. Tail
cancellation improves the situation but in the case of the Exp1TC
method the correlation is not fully removed. However, the
correlations are nicely removed by the Exp2TC and TM+TC methods.

The best results at the described conditions are obtained using the
TM+TC method ($\sigma_y=232\,\mu$m and $\sigma_\phi=0.687\,^\circ$).
With the Exp1TC method we find $\sigma_y=337\,\mu$m and
$\sigma_\phi=0.686\,^\circ$.

For $N_{fit}$ independent fitpoints the relation between the
accuracy of the measurement in $y$ of the single points (here
represented by $\sigma_y$) and of the angular resolution
$\sigma_\phi$ is given by \cite{blum}

\begin{equation}\label{rescomp}
    \sigma_\phi \ \approx \
    \sqrt{\frac{12}{N_{fit}}}\,\frac{\sigma_y}{D} \quad {\bf rad}\ .
\end{equation}

Following Eq. \ref{rescomp}, we expect for about $N_{fit}=20$
fitpoints (see Fig. \ref{event2}), for a detector thickness of
3.7\,cm and for a position resolution of $\sigma_y=337\,\mu$m -- as
measured with the Exp1TC method -- an angular resolution of
$\sigma_\phi=0.4\,^\circ$. For the TM+TC method we expect an even
better angular resolution: $\sigma_\phi=0.23\,^\circ$. The measured
angular resolution does not reach these expected numbers. We
conclude that the fitpoints are not independent as presumed by Eq.
\ref{rescomp}, since the different correction methods supposedly do
not remove fully the correlations between the signals in subsequent
time bins (between the fitpoints).

\subsection{Cluster Reconstruction}
\label{Reconstr}

For each time bin charge sharing between adjacent pads allows to
reconstruct the position of the clusters along a pad row (in the
wire direction). To calculate the $y$-position of that cluster (Fig.
\ref{xsec}) we assume a Gaussian PRF\footnote{A
  simpler method is to calculate the center of gravity on three pads,
  but the PRF method yields results that are more accurate by about
  10\,\%.}. The amplitudes in at least two neighbouring pads are
required to be above threshold, which is determined by the value of
the noise $N$. $N$ was extracted from the baseline in the presamples
of the drift chamber signals. Gaussian fits to the noise
distributions yield values of around 1.7 ADC channels. The
displacement $y_{dis}$ of the cluster from pad $i$ is calculated
using a weighted mean of two measurements \cite{blum}:

\begin{equation}
  y_{dis} \ = \ \frac{1}{w_1+w_2}\left[w_1\left(\frac{\sigma_P^2}{W}
  \ln\frac{A_i}{A_{i-1}}-\frac{W}{2}\right)
  + w_2\left(\frac{\sigma_P^2}{W}
  \ln\frac{A_{i+1}}{A_i}+\frac{W}{2}\right)\right] \ .
\label{clusrec}
\end{equation}

Here $\sigma_P$ is the Gaussian width of the PRF, $W$ is the pad
width and $w_1$, $w_2$ are weights: $w_1 = (A_{i-1})^2$, $w_2 =
(A_{i+1})^2$, with $A_i$ being the amplitude on pad $i$. The error
of the cluster position is given by

\begin{subequations}
  \begin{align}
    \sigma_t \ & = \ 50\,\mbox{ns} & \mbox{in time direction and}
    \label{sigmat} \\
    \sigma_y \ & = \ \sqrt{(\sigma_0)^2 + \frac{2}{A^2}}\quad\mbox{cm}
    & \mbox{along }y\, .\label{sigmay}
  \end{align}
\end{subequations}

Here $A=A_{i-1}+A_i+A_{i+1}$ is the sum of the amplitudes on the
three pads, with $A_{i}>A_{i-1}$ and $A_{i}>A_{i+1}$. The parameter
$\sigma_0 \approx 0.03$\,cm is a specific resolution that is
optimized for the best detector performance. It is of the order of
the residuals $\Delta_y$. We also apply a center of gravity
correction to the time coordinate ($x$ in Fig. \ref{xsec}), by
moving the reconstructed clusters in that coordinate according to
the values of the amplitudes in the neighbouring time bins. Assuming
on a given pad the three amplitudes $A_{t-1}$, $A_{t}$ and $A_{t+1}$
at three subsequent time bins, the cluster at time bin number $t$ is
shifted to

\begin{equation}
  t + \delta t \ = \ t + \delta_0 \
  \frac{-A_{t-1}+A_{t+1}}{A_{t-1}+A_{t}+A_{t+1}}\, .
\end{equation}

\begin{figure}[t]
  \begin{center}
    \includegraphics[width=9cm]{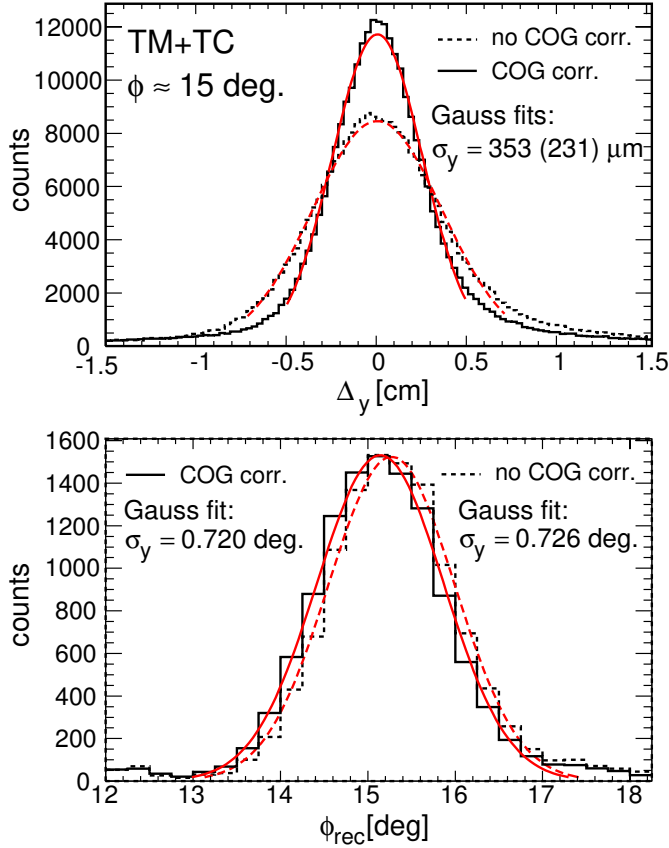}
    \caption{Measured histograms of residuals $\Delta_y$ (top panel)
    and of reconstructed angles (lower panel). We compare data with
    (solid lines) and without (broken lines) time center of gravity
    correction (COG corr.).}
    \label{tcorr}
    \hspace{5mm}
  \end{center}
\end{figure}

$\delta_0 \approx$\ 1.5 is a factor that is optimized for best
detector resolution\footnote{Since now the distance in time
  direction between reconstructed clusters is not constant anymore,
  it seems necessary to change the value of
  the error $\sigma_t$ given by Eq. \ref{sigmat}. However, the
  effect on the resolution obtained is small and we keep $\sigma_t$
  constant, as in Eq. \ref{sigmat}. The result of the time center
  of gravity correction can be seen in Fig. \ref{event2}.}.
This procedure corrects for the ambiguity of the position of the
signal within the 100\,ns time bins and significantly improves the
position resolution: From Fig. \ref{tcorr} we find an improvement of
35\,\%. Despite the better position resolution, the angular
resolution is improved by only 1\,\%. Following Eq. \ref{rescomp}
this would again imply that to some extend the correlations between
the signals in subsequent time bins cannot be removed. The obtained
angular resolutions seem to be the lowest limit.

\subsection{Non-Linearities} \label{sys}

\begin{figure}[t]
  \begin{center}
    \includegraphics[width=9cm]{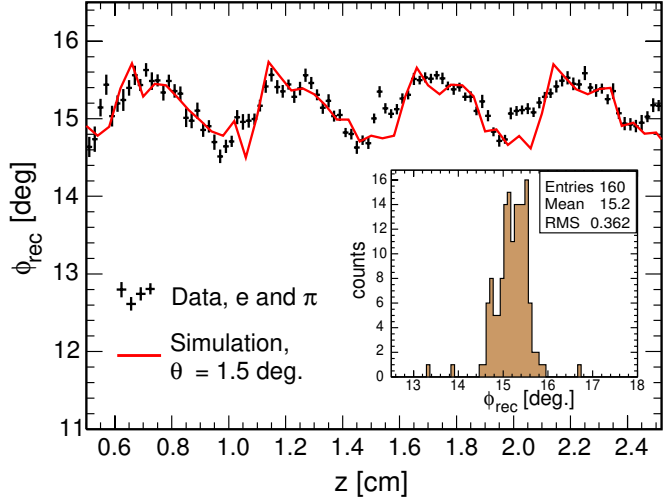}
    \caption{Systematic variation of the reconstructed angle
    $\phi_{rec}$ with the $z$-coordinate (across the wires). We
    show measured (crosses) and simulated results (solid line).
    The insert shows a projection of the measured data on the
    ordinate, giving the overall effect due to non-linearity.
    The TM+TC method was used.}
    \hspace{5mm}
    \label{angz}
  \end{center}
\end{figure}

Fig. \ref{angz} shows the systematic variation of the reconstructed
angle $\phi_{rec}$ with the $z$-coordinate across the wires,
extracted from the silicon strip detectors. Clearly visible is the
influence of the anode wire grid with 0.5\,cm periodicity. This
systematic effect can be approximately reproduced by the simulation
and is explained by the non-linearities in the time-space
relationship as discussed in section \ref{NonIso}. A small variation
of the angle $\theta$ from zero has to be assumed (here
$\theta=1.5\,^\circ$), indicating a slight misalignment of the
chamber with respect to the beam. In fact, the precision of the
alignment in this direction was of this order. If this is the case,
the lateral distance of electrons deposited along the tracks from
the nearest anode wire in the drift cell ($z'$ in Fig. \ref{noniso})
is varying with the distances from the anode wire plane $x$. Thus an
offset that depends on $x$ is added to the drift time of the
electrons, introducing the observed systematic effect. The
resolution deterioration due to this effect is about 0.36\,$^\circ$,
at these specific conditions (pions, 3\,GeV/c, $\phi \approx
15\,^\circ$). For $\theta=0\,^\circ$ the systematic effect
disappears in the simulated data.

\begin{figure}[t]
  \begin{center}
    \includegraphics[width=9cm]{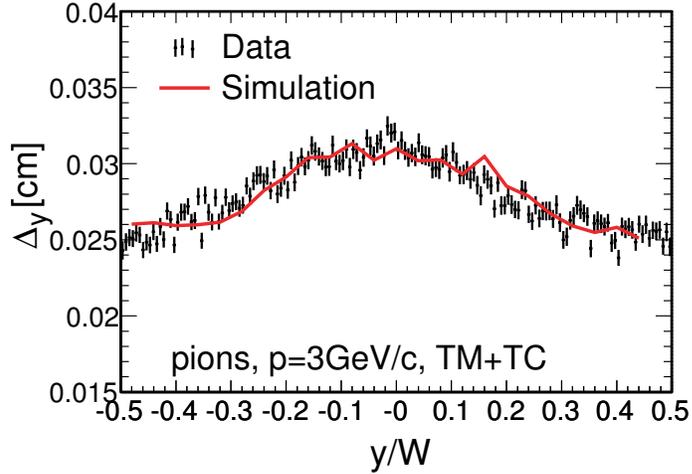}
    \caption{Residuals $\Delta_y$ of the fit as a function of the
    position of the avalanche with respect to the pad. We
    show measured (crosses) and simulated results (solid line).
    The TM+TC method was used. $W$ is the pad width.}
    \hspace{5mm}
    \label{posy}
  \end{center}
\end{figure}

Fig. \ref{posy} shows the residuals $\Delta_y$ as a function of the
position of the avalanche with respect to the pad. This value is a
measure of the error of the coordinate measurement using Eq.
\ref{clusrec}. For avalanches in the center of a pad the error in
the measurement is about 20\,\% larger. Since the same
reconstruction method is used in the simulations, this effect is
also well reproduced.

\section{Position and Angular Resolution}
\label{PosRes}

Here we present the measured detector performance and compare it to
AliRoot simulations.

\subsection{Dependence on $S/N$} \label{PosRes2}

In this section we study the performance of the detector as a
function of the signal-to-noise ratio ($S/N$). The signal height $S$
was extracted from pulse height spectra (mean value) at a given time
bin corresponding to the center of the drift region (1.6\,$\mu$s
drift time at nominal conditions as in Fig. \ref{signals}).

\begin{figure}[t]
  \begin{center}
    \includegraphics[width=9cm]{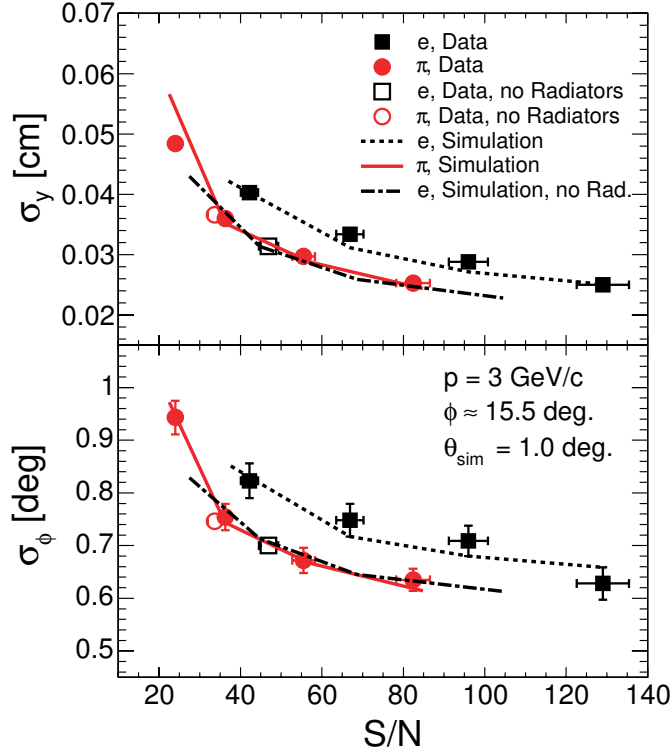}
    \caption{Position resolution $\sigma_y$ and angular resolution
    $\sigma_\phi$ as a function of the signal-to-noise ratio ($S/N$).
    The filled squares (circles) show the measured data for four
    different gain values for electrons (pions) with radiators in
    front of the drift chambers. The large open squares (circles)
    show the measured data for electrons (pions) with no radiators
    and for one given gain. The lines show simulation results for
    different gains for pions and electrons with and without
    radiators.}
    \hspace{5mm}
    \label{resston}
  \end{center}
\end{figure}

The dependence of $\sigma_y$ and $\sigma_\phi$ on $S/N$ is shown in
Fig. \ref{resston}. Again the incident angle of the beam was about
$\phi = 15\,^\circ$. The measured data are nicely reproduced by the
AliRoot simulations for $\theta=1\,^\circ$ (see Fig. \ref{xsec}).
The measured data points for pions and electrons lie on two separate
curves, roughly of $1/\sqrt{S/N}$ form. At a given $S/N$ value, the
resolution is worse for electrons as compared to pions. Since the
$S/N$ value at a given gas gain is about 60\% larger for electrons,
at normal operation conditions the resolution is very similar for
both particle types. The data points without radiators shows better
resolutions for electrons and lies on the same curve as the pion
data\footnote{The $S/N$ value for electrons decreases without
    radiators due to the absence of energy deposit by TR.},
while for pions the performance is similar with and without
radiators. This indicates that the deterioration of the resolution
in the case of electrons with radiators is connected with one of the
following two processes:

\begin{itemize}
\item[i)] Bremsstrahlung created in the radiator,
\item[ii)] Transition Radiation from the radiator.
\end{itemize}

The simulations reproduce the observed behaviour well, implying that
the bremsstrahlung contribution is very small. The processes
secondary to the TR absorption on the other hand turn out to cause a
significant deterioration of the detector resolution for the
electrons. As described in section \ref{TRabsor}, L-shell
fluorescence X-rays are very common. They carry an energy of about
5\,keV and their mean free path is about 0.4\,cm in Xenon. Their
influence on the resolution is dominant; the influence of the range
of photoelectrons and Auger electrons on the other hand is
small\footnote{This agrees well with our observation that a
  magnetic field of 0.14 to 0.56\,T has no influence on the resolution.
  The tracks of photo and Auger electrons would be curled up in
  magnetic fields of that strength, which would lead to an improved
  resolution.}, as well as the influence of K- and M-shell
fluorescence X-rays. The former are high-energetic and generally
escape from the region where the TR absorption takes place. The
latter are low-energetic and their absorption length is too small
($\approx100$\,$\mu$m) to effectively influence the resolution.

\subsection{Dependence on Incident Angle} \label{PosRes2}

\begin{figure}[ht]
  \begin{center}
    \includegraphics[width=12cm]{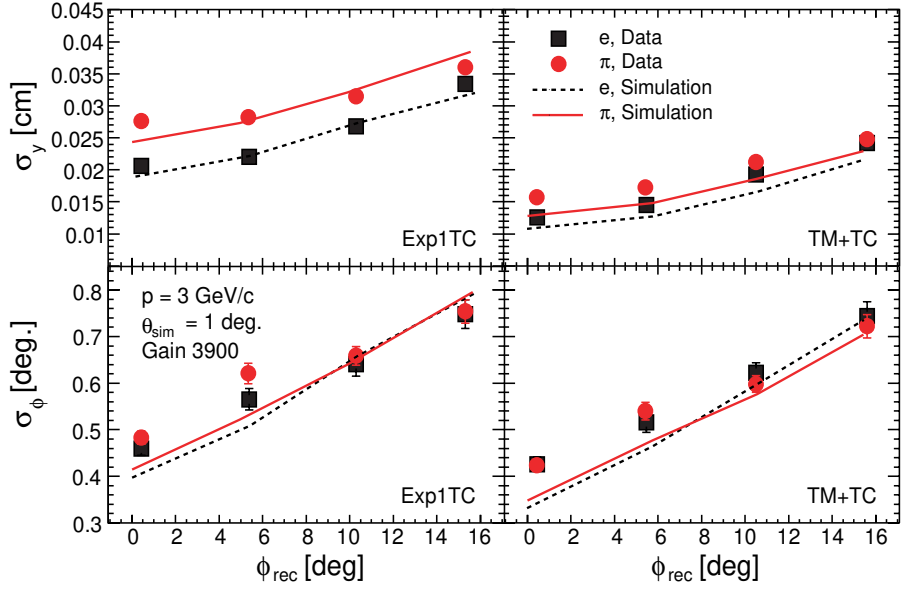}
    \caption{Position resolution $\sigma_y$ (upper panels) and angular
    resolution $\sigma_\phi$ (lower panels) as a function of the
    reconstructed angle for the Exp1TC (left panels) and TM+TC
    (right panels) methods. The radiators were mounted in front of
    the drift chambers and the simulated data correspondingly
    contains TR in the case of electrons.}
    \hspace{5mm}
    \label{resang}
  \end{center}
\end{figure}

Fig. \ref{resang} shows the position resolution $\sigma_y$ and the
angular resolution $\sigma_\phi$ as a function of the reconstructed
angle $\phi_{rec}$ for two tail cancellation methods. The measured
results are quantitatively reproduced by the AliRoot simulations.
Only at small angles is the simulated resolution generally better
than the measurements. This systematic deviation can be explained by
a space charge effect \cite{sc}. At small values of the incident
angle (perpendicular tracks) all electrons created along a track
drift to the same anode wire spot\footnote{If a magnetic field is
applied, the angle $\phi$ where the space charge effect is largest
is modified by the Lorentz angle $\phi_L$ (see Eq. \ref{lorang}).},
leading to a buildup of positive ions around this wire spot. This
reduces the effective gain and thus the $S/N$ value for these
conditions. As a consequence, we observe a deterioration of the
resolution for small angles, which is not reproduced by the
simulations, since no space charge effect is included.

For the Exp1TC method the measured position resolution is around
350\,$\mu$m at $\phi_{rec}\approx 15\,^\circ$ and improves for
smaller values of the incident angle. For the TM+TC method the
position resolution is below 250\,$\mu$m at all investigated angles.
The angular resolution is around $0.7\,^\circ$ at $\phi_{rec}\approx
15\,^\circ$ and below that value for smaller incident angles and
both tail cancellation methods. While the TM+TC method does not
improve the measured angular resolution -- as was already mentioned
-- we observe a slight improvement in the simulated angular
resolution. The observed performance is well within the requirements
for the ALICE TRD that were listed in section \ref{Intro}.

\subsection{Dependence on Drift Velocity} \label{PosDrift}

\begin{figure}[t]
  \begin{center}
    \includegraphics[width=9cm]{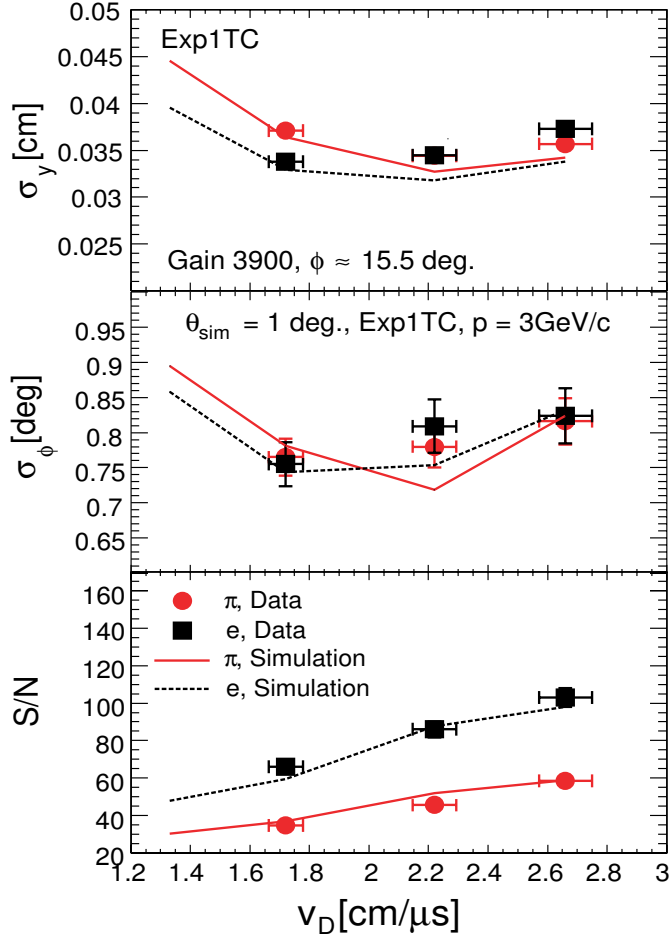}
    \caption{Measured and simulated dependence of the detector
    performance on the drift velocity for electrons and pions.
    From top to bottom the position resolution $\sigma_y$,
    angular resolution $\sigma_\phi$ and signal-to-noise ratio
    $S/N$ are shown. The drift velocity scan was carried out with
    radiators.}
    \hspace{5mm}
    \label{resvd}
  \end{center}
\end{figure}

The nominal drift velocity for the ALICE TRD of around
$v_0=1.5$\,cm/$\mu$s in the drift region was determined together
with the sampling rate of 10\,MHz to lead to a sufficiently large
number of fit points for the track reconstruction ($N_{fit}\approx
20$, see Fig. \ref{event2}). However, to provide more general
results we also varied $v_D$ and investigated its influence on the
resolution of the detector. Fig. \ref{resvd} shows the dependence of
the position resolution $\sigma_y$ and of the angular resolution
$\sigma_\phi$ on $v_D$. In the measurements the average drift
velocity $v_D^{av}\approx v_D$ can be extracted from the data using
Eq. \ref{trackvd}, if the incident angle of the beam is known.
Values for $v_D$ were varied by setting the drift voltage to 2.1
(nominal value), 2.4 and 2.7\,kV, while keeping the anode wire
voltage constant at 1.55\,kV, corresponding to a gain of 3900. The
two main contributions to the resulting resolution values are the
lever arm of the fit (number of fit points) and the average
amplitude per time bin. The number of fit points depends inversely
on the drift velocity. The average amplitude per time bin is
increased by a larger drift velocity, since more electrons reach the
anode wires per time unit, leading to an increase of the $S/N$
value. As a consequence we find, in Fig. \ref{resvd}, that a large
drift velocity leads to a deterioration in the resolution due to the
decrease in the number of fit points. Accordingly, the resolution
also deteriorates for small drift velocities, due to the decrease of
$S/N$. Our nominal conditions in the beam test -- namely a drift
voltage of 2.1\,kV, corresponding to a drift velocity of around
$v_D=1.67$\,cm/$\mu$s -- turn out to be a good choice for our
specific chamber dimensions and readout rate (10\,MHz).

\subsection{Dependence on Momentum} \label{PosMom}

\begin{figure}[th]
  \begin{center}
    \includegraphics[width=9cm]{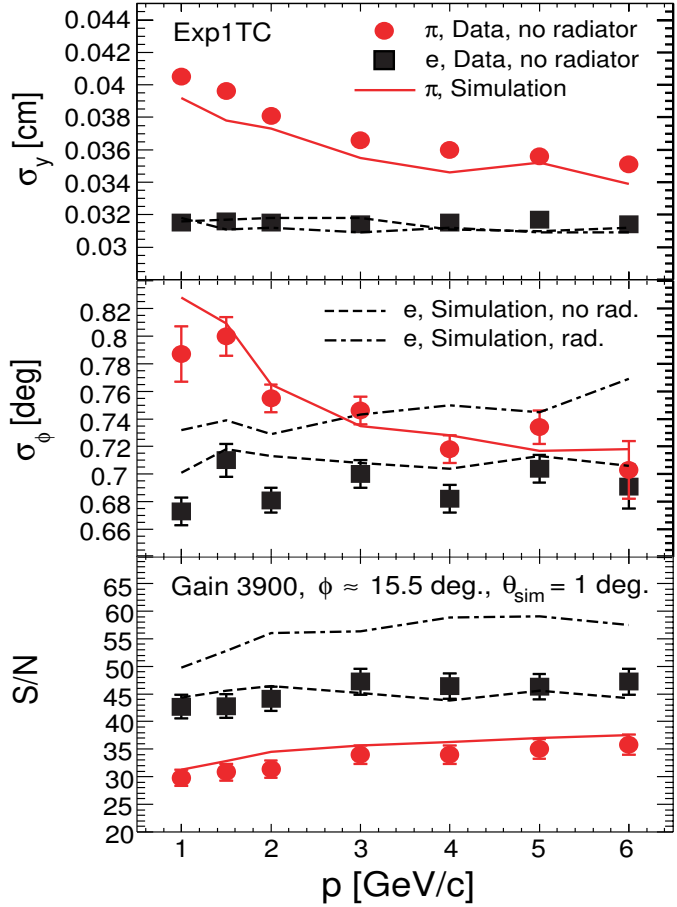}
    \caption{Measured and simulated dependence of the detector
    performance on the particle momentum for electrons and pions.
    The position resolution $\sigma_y$, angular resolution
    $\sigma_\phi$ and signal-to-noise ratio $S/N$ are shown (from
    top to bottom). The experimental momentum scan was carried out
    without radiators. However, here we added also simulated data
    with TR to demonstrate its momentum-dependent effect on the
    resolution of the electrons.}
    \hspace{5mm}
    \label{resmom}
  \end{center}
\end{figure}

In  Fig. \ref{resmom} we show the dependence of the resolutions
$\sigma_y$ and $\sigma_\phi$ on the beam momentum $p$. Generally,
$\sigma_y$ and $\sigma_\phi$ for pions improve for larger momenta,
which is explained by the increased $S/N$ value for larger momenta.
The ratio $S/N$ as a function of $p$ is shown in the lower panel of
Fig. \ref{resmom}. At $p\approx 0.56$\,GeV/c, corresponding to
$\beta\gamma=p/(m_\pi\,c)\approx 4$ (where $m_\pi\approx 140$\,MeV is
the mass of the pion), pions are minimum ionizing; at larger momenta
-- especially in the momentum range of interest between 1 and
6\,GeV/c -- the ionization energy loss and thus the measured $S/N$
value continuously increase \cite{dedx}. Electrons already have a
factor $\beta\gamma\approx 2000$ at $p=1$\,GeV/c momentum. Thus they
are at the plateau of ionization energy loss. In the momentum range
of interest, the $S/N$ value is constant (with no radiators),
leading to a constant resolution as a function of the momentum. The
situation is changed with radiators, since at an electron momentum
of $p = 1$\,GeV/c the TR production sets in
and at higher momenta the energy deposit due to TR and thus the
$S/N$ value increase considerably. However, due to the effects
described in section \ref{PosRes2} the performance is not improved
by the larger energy deposit associated with TR, but deteriorated.

\subsection{Performance in Magnetic Field}
\label{Bfield}

\begin{figure}[t]
  \begin{center}
    \includegraphics[width=9cm]{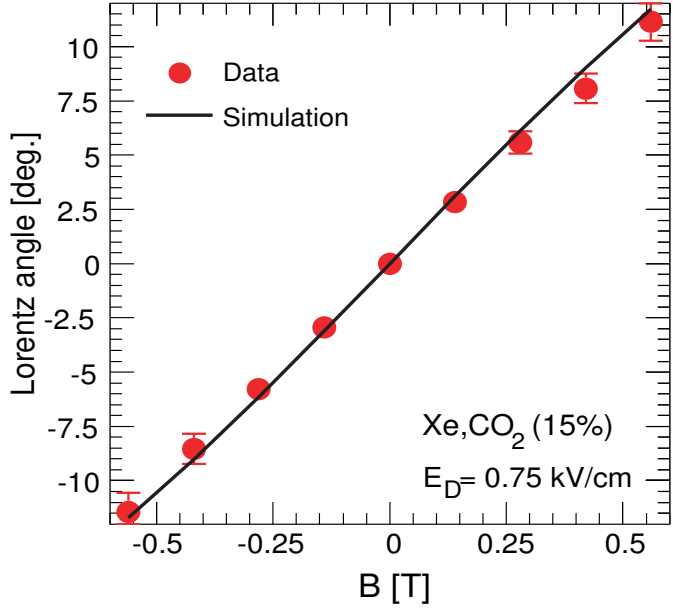}
    \caption{Measured and calculated Lorentz angle for the Xe, CO$_2$
    mixture as a function of the magnetic field $B$. The calculations
    were carried out with the MAGBOLTZ program. The drift field
    strength is $E_D=0.75$\,kV/cm.}
    \hspace{5mm}
    \label{lorentz}
  \end{center}
\end{figure}

The ALICE TRD will be situated inside the large L3 barrel magnet
with a magnetic field of 0.4\,T. Since the electrons drift
perpendicularly to the field, they will experience the Lorentz force
that leads to a displacement of the clusters along the pad rows as a
function of the drift time. For an electron produced at position
($x_0$, $y_0$, $z_0$), where the $y$ and $z$ directions are
perpendicular to the drift direction along $x$ and the $y$ direction
is parallel to the wires (see Fig. \ref{xsec}), the new $y$ position
is given by

\begin{equation}\label{lorang}
    y \ = \ y_0 \ + \ \omega\,\tau\,(x-x_0),\qquad
    \omega\tau \ = \ \tan \phi_L \ ,
\end{equation}

where $\phi_L$ is the Lorentz angle. It is visible as an apparent
inclination of the reconstructed track. If the track passing through
the detector has an angle $\phi_0$, then the reconstructed angle
$\phi_{rec}$ is given by

\begin{equation}\label{lorang2}
    \phi_{rec} \ = \ \phi_0 \ - \ \phi_L \ + \ \delta \phi \ ,
\end{equation}

where $\delta\phi$ is the error of the measurement. The Lorentz
angle depends on the magnetic field strength and the drift velocity
of the electrons. This dependence needs to be known to be able to
reconstruct the original particle track and extract its inclination
$\phi_0$.

We measured the Lorentz angle as a function of the magnetic field
and of the electron drift velocity and compare the results to
MAGBOLTZ \cite{magboltz} calculations (Fig. \ref{lorentz}). For a
magnetic field of 0.4\,T and a drift field of 750\,V/cm we find a
Lorentz angle close to 8$\,^\circ$.

\begin{figure}[t]
  \begin{center}
    \includegraphics[width=9cm]{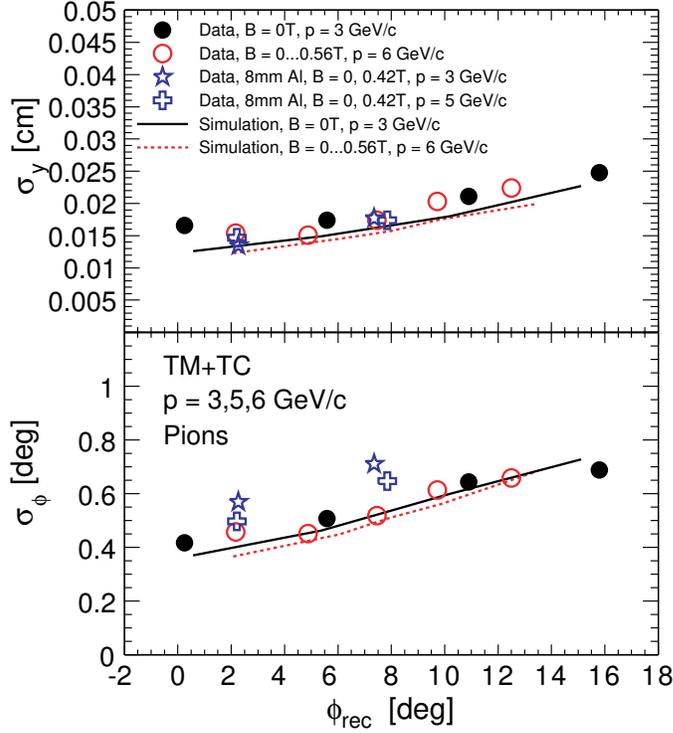}
    \caption{Position resolution $\sigma_y$ (top panel) and angular
    resolution $\sigma_\phi$ (lower panel) as a function of the
    reconstructed angle with and without magnetic field for pions.
    The symbols show the measurements, the lines show simulated data.
    In the data with magnetic field the actual incident beam angle
    was around $\phi_0 = 2.2^\circ$. The magnetic field dependent
    Lorentz angle is added to $\phi_0$ following Eq. \ref{lorang2}.
    We also show measured data taken with a 0.8\,cm thick aluminum
    plate in front of the detectors (stars and crosses).}
    \hspace{5mm}
    \label{resfield}
  \end{center}
\end{figure}

In Fig. \ref{resfield} we show the dependence of the resolution on
the Lorentz angle $\phi_L$. It is very similar to the dependence on
$\phi_0$, which implies that the resolution depends only on the
value of reconstructed angle $\phi_{rec}$, and that additional
effects due to the presence of the magnetic field are negligible. In
Fig. \ref{resfield} we also show data taken with a 0.8\,cm thick
aluminum plate in front of the detectors. This plate has a radiation
length of about $0.9 X_0$, similar to that of four TRD layers. As
expected, the resolution in $\sigma_y$ is undisturbed, since it does
not depend on external effects like multiple scattering. On the
other hand, we find clear effects on $\sigma_\phi$, evidencing a
momentum dependent multiple scattering.

\subsection{Performance compared to External Track Reference}
\label{SiPerf}

\begin{figure}[th]
  \begin{center}
    \includegraphics[width=9cm]{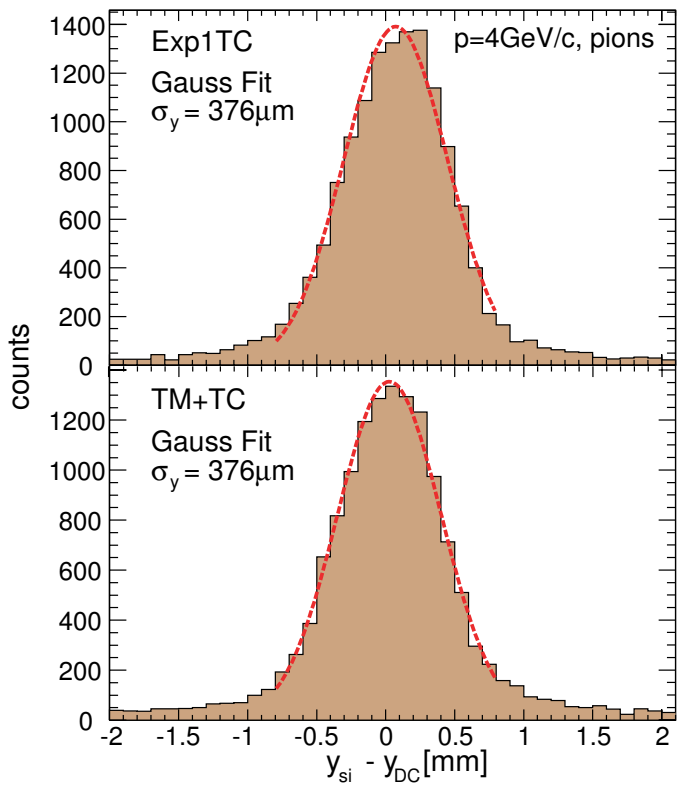}
    \caption{Position resolution of the detectors with respect to
    a silicon strip detector for the Exp1TC (upper panel) and
    TM+TC (lower panel) methods for pions at 4\,GeV/c. The
    extracted position resolution of 376\,$\mu$m for both
    methods is for the reconstructed tracks at the
    center of the detector (time bin 15, 1.5\,$\mu$s drift time, see Fig.
    \ref{event2}). It includes the resolution of the silicon strip
    telescope ($\approx 50\,\mu$m) as well as the beam divergence
    ($\approx 0.1\,^\circ$).}
    \hspace{5mm}
    \label{poshisto}
  \end{center}
\end{figure}

Finally we show in Fig. \ref{poshisto} the position resolution of
the detector with respect to a silicon strip detector, which is
located a few centimeters in front of the investigated chamber. Let
$y_{DC}$ be the position of the center of the reconstructed track
(at time bin 15, see Fig. \ref{event2}) and $y_{si}$ the position
given by the silicon telescope. Then $y_{si}-y_{DC}$ is a measure of
the total position resolution of the drift chamber. This does
however include the resolution of the silicon strip detector
($\approx 50\,\mu$m) as well as the beam divergence ($\approx
0.1\,^\circ$) and external scattering effects. Fig. \ref{poshisto}
shows a histogram of $y_{si}-y_{DC}$ for pions at $p=4$\,GeV/c, at
$\phi_0\approx 15\,^\circ$, and for two tail cancellation methods. A
Gaussian fit yields a position resolutions of 376\,$\mu$m for both
methods, which is similar to the resolution extracted from the
residuals for the same conditions.

\section{Summary and Conclusions}

We have measured the performance of drift chamber prototypes for the
ALICE Transition Radiation Detector (TRD) with respect to position
and angular resolution. The detectors are drift chambers with
cathode pad readout filled with the Xe,CO$_2$(15\%) mixture. For
incident particle angles from 0 to 15$\,^\circ$ with respect to the
wire normal we find a position resolution better than 300\,$\mu$m
($\sigma$) and an angular resolution below 0.8$\,^\circ$ ($\sigma$).
A systematic effect of about 0.36$\,^\circ$ at $\phi=15\,^\circ$ is
introduced by non-linearities: The discrete configuration of the
wire grids in connection with the generally higher drift velocity in
the amplification region introduces a modulation in the electron
drift times, leading to a distortion of the space-time relation
(non-linearity).

If a radiator is added to the drift chambers, transition radiation
contributes to the energy deposit in the gas. Then the
signal-to-noise ratio is increased for electrons, but, nevertheless,
the electron resolution is by about 7\,\% worse in that case.
L-shell fluorescence photons, which are produced in secondary
processes after the absorption of the transition radiation photons,
have an absorption length of about 0.4\,cm in the xenon gas mixture.
This smearing of the charge deposit around the actual track of the
incident electrons introduces a considerable degradation of the
position reconstruction performance for electrons.

The measurements are compared to simulations carried out with
AliRoot, the ALICE event simulation and analysis framework. The
non-linearity of the electron drift was calculated with GARFIELD and
included in the AliRoot code. Also a simplified picture of the
secondary processes following the transition radiation absorption
was added to AliRoot. The charge sharing between adjacent pads (pad
response function) was calculated using an exact method (weighting
field formalism). The performance of the detector is well understood
and the position and angular resolution are within the requirements
for the ALICE TRD. Our results -- in particular the investigated
systematic effects, the corrections applied, and the influence of
the transition radiation -- are of general interest also for other
TRD's and/or other drift chambers with similar geometry, where a
drift region is added to a multiwire proportional chamber, with the
electron drift perpendicular to the wire planes.

\section*{Acknowledgements}

We acknowledge A. Radu and J. Hehner for their skills and dedication
in building our detectors and N. Kurz for his help on data
acquisition. We would also like to acknowledge A. Przybyla for his
technical assistance during the measurements.



\end{document}